\documentclass[aps,pre,twocolumn,preprint,a4paper,10pt,floats]{revtex4} 
\usepackage{graphicx}
\usepackage{float}
\usepackage{epsfig}
\usepackage{verbatim}
\usepackage{color}
\usepackage{grffile}
 
\begin{document} 

\title{Anticipating Persistent Infection}
\author{Promit Moitra$^1$, Kanishk Jain$^2$ and Sudeshna Sinha$^1$}
\email{e-mail: promitmoitra@iisermohali.ac.in}
\affiliation{$^1$Indian Institute of Science Education and Research Mohali, Sector 81, PO Manauli 140306, Punjab, India\\$^2$Department of Physics, Emory University, Atlanta GA 30322}

\begin{abstract} 
We explore the emergence of persistent infection in a closed region where the disease progression of the individuals is given by the SIRS model, with an individual becoming
infected on contact with another infected individual within a given range. We focus on the role of synchronization in the persistence of contagion. Our key result is that higher degree of synchronization, both globally in the population and locally in the neighborhoods, hinders persistence of infection. Importantly, we find that early short-time asynchrony appears to be a consistent precursor to future persistence of infection, and can potentially provide valuable early warnings for sustained contagion in a population patch. Thus transient synchronization can help anticipate the long-term persistence of infection. Further we demonstrate that when the range of influence of an infected individual is wider, one obtains lower persistent infection. This counter-intuitive observation can also be understood through the relation of synchronization to infection burn-out.
\end{abstract}

\maketitle

The spread of infectious diseases in a population is a field of wide-spread inquiry and continues to attract intense research activity \cite{McEvedy}. One of the outstanding problems in this area has been obtaining reliable early warning signals for persistence of infection in a region. This is a problem of obvious significance, as it can potentially influence strategies of long-term control of disease. Mathematically this is a challenging problem, as one has to consider large interactive complex systems that are strongly nonlinear and typically not well mixed. In this work we attempt to uncover what dynamical features at early times are strongly correlated to long-term characteristics, such as the continued presence of infection in a population patch. Such features, if found to exist, can potentially provide important {\em early warning signals} for persistent infections.

Mathematically, epidemiological models have successfully captured the dynamics of infectious disease \cite{murray,sw}. One well known model for non-fatal communicable disease progression is the SIRS cycle. This model appropriately describes the progression of diseases such as typhoid fever, tetanus, cholera, small pox, tuberculosis and influenza \cite{hethcote,ozcaglar}. The SIRS cycle is described by the following stages:
\begin{itemize}
\item 
Susceptible (denoted by symbol $S$) - An individual in this state remains susceptible until they contract the infection from another infected person in their neighbourhood. At the end of the refractory stage (namely the stage of temporary immunity) of the disease cycle, the individuals return to this state.
\item 
Infected (denoted by $I$) - In this stage of the disease cycle, the individual is in an infected state, which signifies they can infect others around them.
\item 
Refractory (denoted by $R$) - At the end of the infectious stage, the individuals acquire temporary immunity to the disease. In this stage they neither get infected by infectious neighbors, nor do they infect anyone in their surroundings. Typically, this stage lasts longer than the infected stage, and at the end of this stage the individual loses the temporarily acquired immunity and becomes susceptible to the infection once again.
\end{itemize}

So the progression of an individual from the susceptible stage, to the infected stage, onto the refractory stage and back to susceptible ($S \rightarrow I \rightarrow R \rightarrow S$) is the SIRS {\em disease cycle}. Cellular automata models \cite{ermentrout} of this cyclic disease progression have provided very good test-beds for studying infection spreading \cite{kuperman,gade,kohar}. In this class of models we consider individuals located on a plane, namely each individual is indexed on a $2$ dimensional lattice by a pair of site indices $(i,j)$. The state of each individual is characterized by a integer-valued counter $\tau_{i,j} (t)$ that describes its {\em phase} in the cycle of the disease, at discrete time step $t$ \cite{kuperman}. Here $\tau$ can take  values $ 0, 1 \dots \tau_I, \tau_I+1 \dots , \tau_I + \tau_R$. At any instant of time $t$ (where $t$ is integer-valued), if phase $\tau_{i,j}$(t) = $0$, then the individual at site $(i,j)$ is susceptible ($S$); if $1 \le \tau_{i,j}(t) \le \tau_I$, then it is infected ($I$); if phase $\tau_{i,j} (t) > \tau_I$, it is in the refractory stage ($R$). 

The dynamics is given as follows: for infected or refractory individuals whose phase $\tau_{i,j} (t) \ne 0$, $\tau_{i,j}$ increases by $1$ at the subsequent time step. Additionally, at the end of the refractory period i.e. when namely $\tau = \tau_0$, where $\tau_0=\tau_I+\tau_R$, the individual becomes susceptible again (characterized by $\tau = 0$). So this implies that if $\tau_{i,j} (t) = \tau_0$, then $\tau_{i,j} (t+1) = 0$. The complete set of evolution rules can then be summarized mathematically as:
\begin{eqnarray}
\tau_{i,j} (t+1)&=&\tau_{i,j} (t) + 1 \hfill  \rm{ if } \ 1 \leq \tau_{i,j} (t) < \tau_0 \\
 &=&0 \hfill  \rm{\ if } \ \tau_{i,j} (t) = \tau_0
\label{sirs}
\end{eqnarray}
The total length of the disease cycle, denoted by $\tau_D$, is equal to $\tau_I + \tau_R + 1$, including the state $\tau = 0$ the individual returns to at the end of the refractory period.
In this work we consider the typical condition where the refractory stage is longer than the infective stage, i.e. $\tau_R > \tau_I$.

{\em Spatiotemporal evolution of infection:}
We now investigate the spread of disease in a spatially distributed group of individuals, where at the individual level the disease progresses in accordance with the SIRS cycle described by the Cellular Automaton model above. In particular, we consider a population of individuals on a $2$-dimensional square lattice of linear dimension $L$, where every node represents an individual \cite{rhodes}. Unlike many earlier studies, we are interested in a {\em closed} patch of individuals. So instead of the commonly used periodic boundary conditions, the boundaries of our system are fixed, with no individuals outside the boundaries. We will focus on the emergence of persistent infection in such an isolated patch.

We consider a following condition for spread of infection: if one or more of its nearest neighbours of a susceptible individual is infected, then the susceptible individual will become infected. That is, if $\tau_{i,j} (t) = 0$, (namely, the individual is susceptible), then $\tau_{i,j} (t+1) = 1$, if any $1 \le \tau_{x,y} (t) \le \tau_I$ where $x,y$ belong to a neighbourhood consisting of $4$ individuals, namely the von Neumann neighborhood, given by: $x,y \in \left\{(i-1,j),(i,j+1),(i+1,j),(i,j+1)\right\}$. Further, we will also consider a neighbourhood comprising of $8$ individuals, namely the Moore neighbourhood, given by: $x,y \in \{(i-1,j),(i,j+1),(i+1,j),(i,j+1),(i-1,j-1),$$ $$(i+1,j+1),(i-1,j+1),(i+1,j-1)\}$ We denote the number of neighbours by $K$, with the von Neumann neighbourhood having $K=4$, while the Moore neighbourhood has $K=8$. Larger $K$ implies that an infected individual can affect individuals in a larger zone around it, namely the infected individual has a larger range of influence. So the dynamics of this extended system combines deterministic, as well as probabilistic elements. The disease progression of an infected individual is deterministic, with the infected period of length $\tau_I$, followed by the refractory period of length $\tau_R$. However, the process of contracting the infection is probabilistic, arising from the interplay of the localized nature of the interactions and the random initial states of the individuals. 

Now the infection in this closed patch can either die out, or it can persist. So it is of considerable significance to find the conditions that lead to sustained infection, as well as to uncover the salient features that characterize the persistent state. The relevant quantity here is the asymptotic fraction of infected individuals in the population. To obtain an appropriate measure of this we first find the fraction of infecteds at time $t$, denoted by $I(t)$. In order to gauge asymptotic trends, we consider this fraction of infected individuals, after long transient time, averaged over several disease cycles, denoted by $\langle I \rangle$. This quantity serves as an order parameter for persistent infection, with non-zero $\langle I \rangle$ indicating persistent infection, while $\langle I \rangle = 0$ indicates that infection has died out in the patch. Further we consider the ensemble averaged $\langle I \rangle$, denoted by $\langle \langle I \rangle \rangle$. This quantity reflects the the size of the basin of attraction of the persistent state, and indicates the probability of persistent infection arising from a generic random initial condition of the population. So $\langle \langle I \rangle \rangle$ is non-zero when persistent infection arises from typical initial states and zero otherwise. 

By studying the dependence of $\langle \langle I \rangle \rangle$ on the initial fraction of infecteds $I_0$, susceptibles $S_0$ and refractory individuals $R_0$ in the population, it was found in Ref.~\cite{scirep} that for sustained contagion in a population, the initial population needed to be a well mixed heterogeneous collection of individuals, with sufficiently large number of both susceptible and refractory individuals. Further, it was found that in a population composed of an admixture of susceptible and refractory individuals, persistent infection emerged in a window of reasonably low $I_0$, with $I_0 \rightarrow 0$ at the lower end of this persistence window. Similar phenomena have been observed in diseases modeled by SEIR  where spatial heterogeneity was seen to play an important role in the persistence of disease \cite{lloyd}. Further, in more general terms, these results are reminiscent of the observation of noise-sustained oscillations of excitable media \cite{kurths}, with noise playing the role of heterogeneity.

In this work we will focus on the {\em correlation between synchronization and persistent infection}. We ask two complementary questions: First, does lack of synchronization characterize the state of the population where infection is sustained. Secondly, and more significantly, does the lack of synchronization in the {\em early stages of disease spreading} lead to persistent infection at later times. We will explore this question by {\em introducing local and global measures of synchronization}. Lastly, we will demonstrate that when the range of infection transmission of an infected individual is wider, one obtains lower persistent infection. We will account for this counter-intuitive observation through the relation between synchronization and infection burn-out.

\noindent  
{\bf Synchronization characterizes populations with sustained infection}:
We first explore the degree of global synchronization in the system, by calculating the quantity:
\begin{equation}
\label{eqn:sigma}
\sigma(t) = | \frac{1}{N} \Sigma^N \exp^{i \phi_{m,n} (t)} |
\label{sigma}
\end{equation}
where $\phi_{m,n} = 2 \pi \tau_{m,n}/\tau_D$ is a geometrical phase corresponding to the disease stage $\tau_{m,n}$ of the individual at site $(m,n)$. Here the indices $m$ and $n$ run from $1$ to $L$, namely over all $N = L \times L$ individuals in the population patch. We use Eqn.~\ref{sigma} to obtain the asymptotic time averaged synchronization order parameter, denoted by $\langle \sigma \rangle$, by averaging $\sigma(t)$ over time, of the order of several disease cycles, after transience. This reflects the synchronization in the emergent system, namely the asymptotic degree of synchronization in the population arising from a specific initial state. So when the phases of disease of the individuals are uncorrelated, that is the disease cycles of the individuals in the population are not synchronized, $\langle \sigma \rangle$ is close to $0$. On the other hand when the individual disease cycles are quite synchronized, $\langle \sigma \rangle$ tends to $1$. We will then go on to calculate the ensemble averaged asymptotic synchronization order parameter denoted by $\langle \langle \sigma \rangle \rangle$, obtained by further averaging the time-averaged asymptotic synchronization order parameter $\langle \sigma \rangle$ over a large number of initial states characterized by a specific ($I_0, S_0, R_0$). This order parameter indicates the probability of synchronization arising from a generic random initial condition of the population. We will use this measure, alongside the ensemble averaged persistence order parameter $\langle \langle I \rangle \rangle$, to help us gauge the broad correlation between synchronization in the emergent population (or lack thereof) with persistent infection.
    
\begin{figure}[H]
\centering
\includegraphics[width=0.8\linewidth,height=0.475\linewidth]{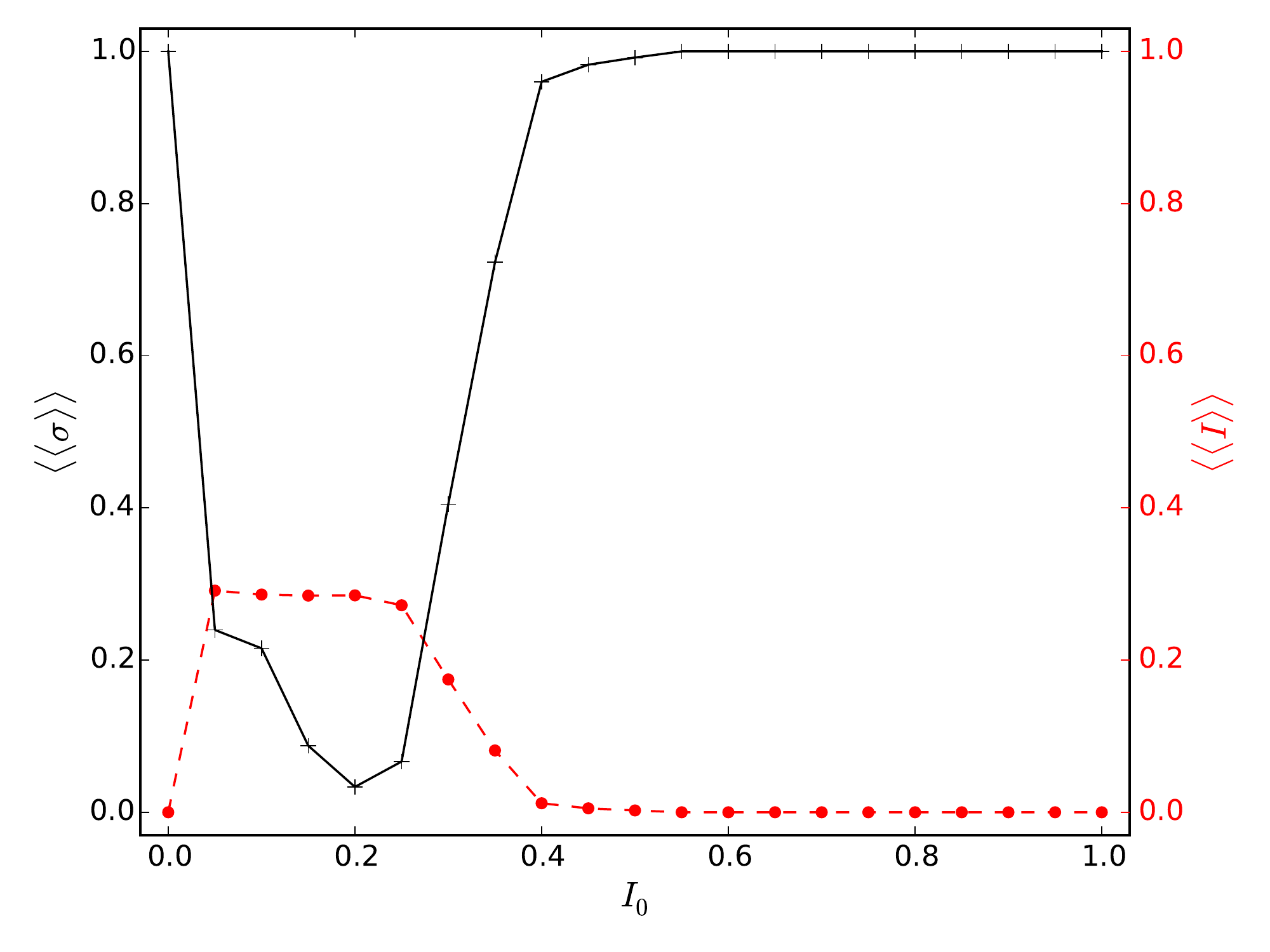}
\caption{\scriptsize{Dependence of the ensemble averaged asymptotic synchronization order parameter $\langle \langle \sigma \rangle \rangle$ (black solid line) on the initial fraction of infecteds in the population $I_0$ (with equal initial fractions of susceptible and refractory individuals: $S_0=R_0$).
Here system size is $100 \times 100$ and $K=4$. The figure also shows the variation of $\langle \langle I \rangle \rangle$ (red dashed line) with respect to $I_0$, where $\langle \langle I \rangle \rangle$ is an ensemble averaged order parameter reflecting the degree of persistence of infection in the population. 
}}
        \label{fig1_k=4}
        \end{figure}
        
    Fig.~\ref{fig1_k=4} shows $\langle \langle I \rangle \rangle$ and $\langle \langle \sigma \rangle \rangle$, for different initial fraction of infecteds $I_0$ in the population. As mentioned before, one observes persistent infection (i.e. $\langle \langle I \rangle \rangle \ne 0$), in a window of $I_0$ \cite{scirep}. Further, it is now clearly evident that in this same window of persistent infection, the global asymptotic synchronization order parameter is the lowest. So {\em higher persistence of infection is consistently correlated with lower degree of synchronization}, distinctly implying that a population where infection is persistent is generally characterized by low synchronization among the individuals. 
Specifically, for instance for the case of persistent infection with $\langle \langle I \rangle \rangle \sim \frac{1}{3}$, we find $0 < \langle \langle \sigma \rangle \rangle < \frac{1}{3}$. On the other hand, for cases where the infection eventually dies out, i.e. $\langle \langle I \rangle \rangle \sim 0$, we have $\langle \langle \sigma \rangle \rangle \sim 1$.  So it is evident that there is clear inverse dependence of infection persistence as reflected by $\langle \langle I \rangle \rangle$ and degree of synchronization of the disease cycles of the individuals in the emergent population as reflected by $\langle \langle \sigma \rangle \rangle$. {\em So one can infer that a population where persistent infection emerges, is quite unsynchronized.}\\
    
\noindent
{\bf Transient synchronization results in weaker persistence of infection:}
We have shown above that lack of synchronization is a key feature of populations with sustained infection, and the asymptotic synchronization order parameter $\langle \sigma \rangle$ successfully characterizes populations with different degrees of persistence of infection. This motivates us to explore the second question: is synchronization in the initial (transient) stage, which we will call {\em transient synchronization} here, an indicator of future persistence of infection in the population? 

\begin{figure*}[htb]
\includegraphics[width=0.475\linewidth]{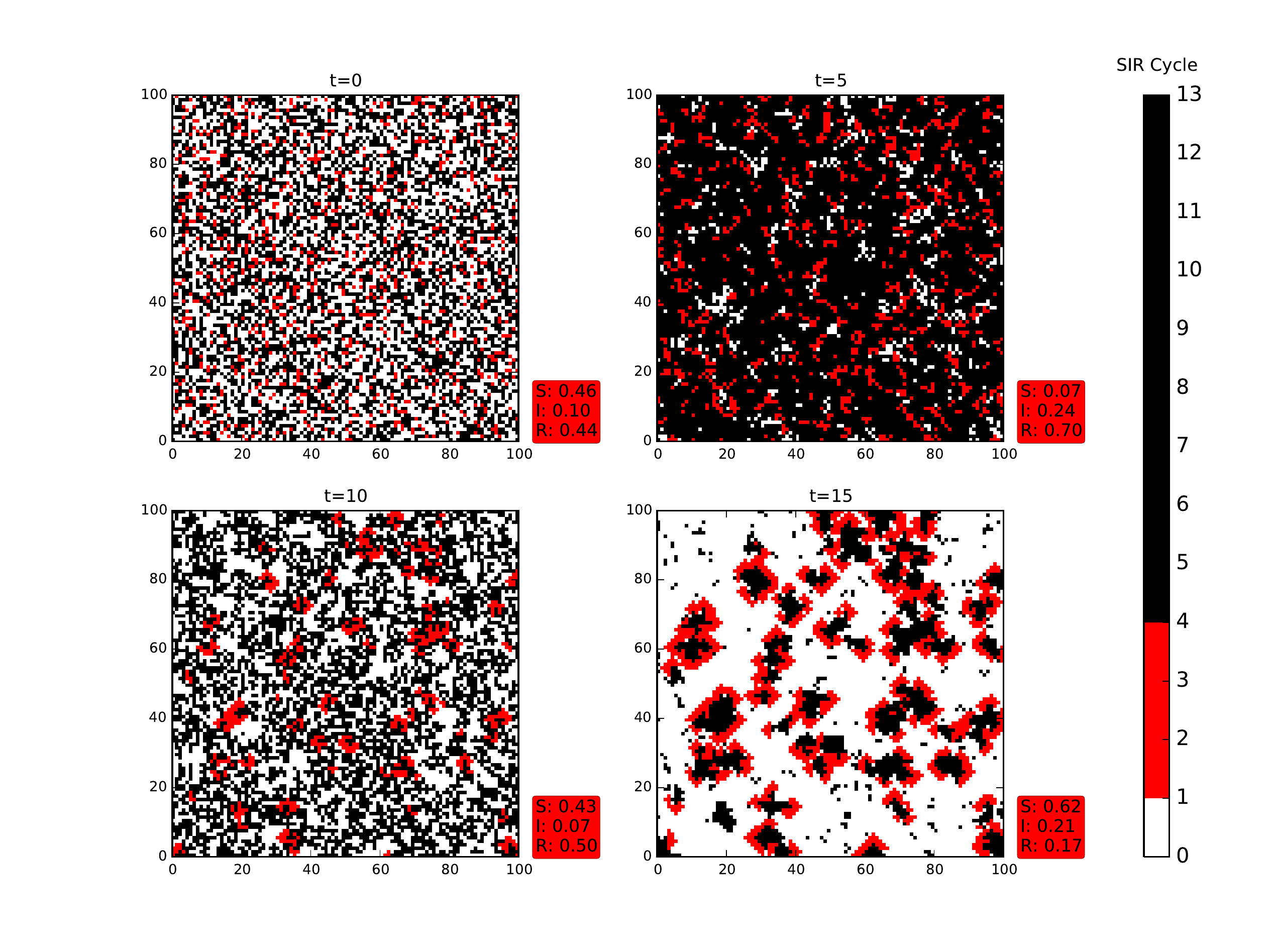} 
\includegraphics[width=0.475\linewidth]{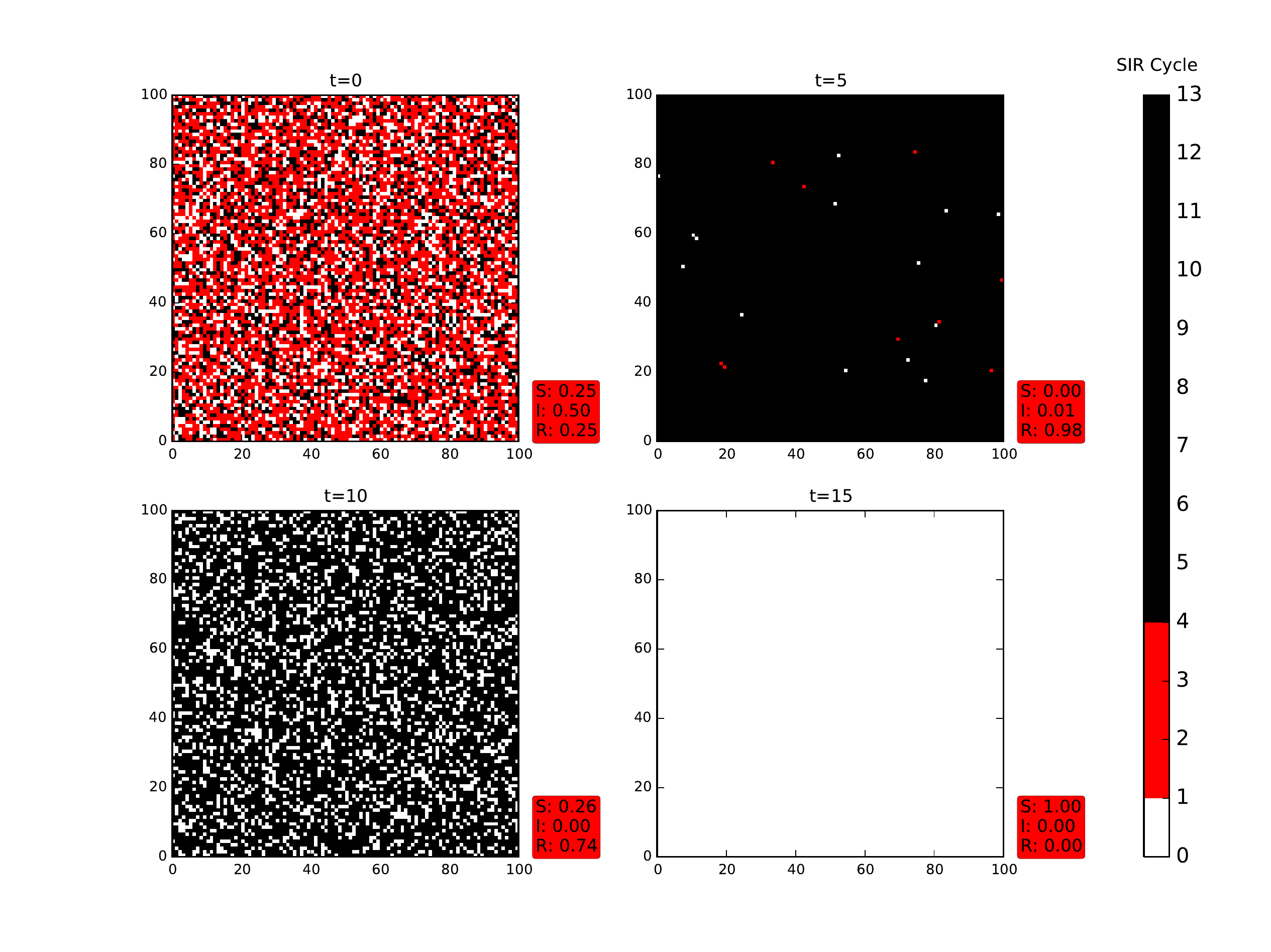} 

\hspace{4cm} \scriptsize{(a)} \hfill \scriptsize{(b)} \hspace{4cm}
\caption{\scriptsize{
Snapshots of the infection spreading pattern at very early times $t=0,5,10,15$, in an initial population comprising of a random admixture of individuals, with $S_0 = R_0$ and (a) $I_0=0.1$ and (b)  $I_0=0.5$. The colour bar shows the relative lengths of the susceptible (S), infected (I) and refractory (R) stages in the disease cycle, where $\tau_I=4$, $\tau_R=9$ and the total disease cycle $\tau_D$ is $14$. The red box shows the fraction of S, I and R individuals in the population at that instant of time. Notice that the population appears to lack of synchrony in the individual states, and has a non-uniform distribution. The infection persists in (a) and dies out in (b).}}
\label{initsync1}
\end{figure*}

First, we show in Figs.~\ref{initsync1}a-b illustrative examples in order to visually examine the state of the system at various instances of time within the first disease cycle, arising from two distinct initial conditions. The first example is a population with initial fraction of infecteds $I_0=0.1$ for which the infection persists, and the second example has $I_0=0.5$ for which the infection burns out (cf. Fig.~\ref{fig1_k=4}). Clearly the case with long-term persistence of infection is marked by a lack of synchrony. On the other hand, the case where the infection dies out shows pronounced synchrony within the {\em first few time steps}. We would now like to investigate if this qualitative observation holds consistently, quantitatively, over a large range of initial states.

{\bf Quantifying finite-time transient synchronization:}
In order to quantify the early time synchronization in the system, we introduce a finite time average of the synchronization order parameter $\sigma (t)$, from the initial time ($t=0$) up to a specific time $t=T$ denoted by $\langle \sigma_T \rangle$. Such a measure reflects the degree of synchronization over short time-scales, at early times. We further consider the ensemble average of this quantity, where the average is over a large set of initial states with a specific initial partitioning ($I_0, S_0, R_0$). This quantity reflects the degree of synchronization typically arising up to time $T$ in the population, from a generic initial state, for a specific ($I_0, S_0, R_0$), and is denoted by $\langle \langle \sigma_T \rangle \rangle$. When $T \sim \tau_0$ (of the order of a single disease cycle), this quantity reflects the transient synchronization or early-time synchronization, namely synchronization of the population within the first cycle of disease. In this work we will aim to {\em explore if this quantity can offer a consistent early warning signal for persistence of infection in the patch of population.}

    \begin{figure}[htb]
\includegraphics[width=0.475\linewidth]{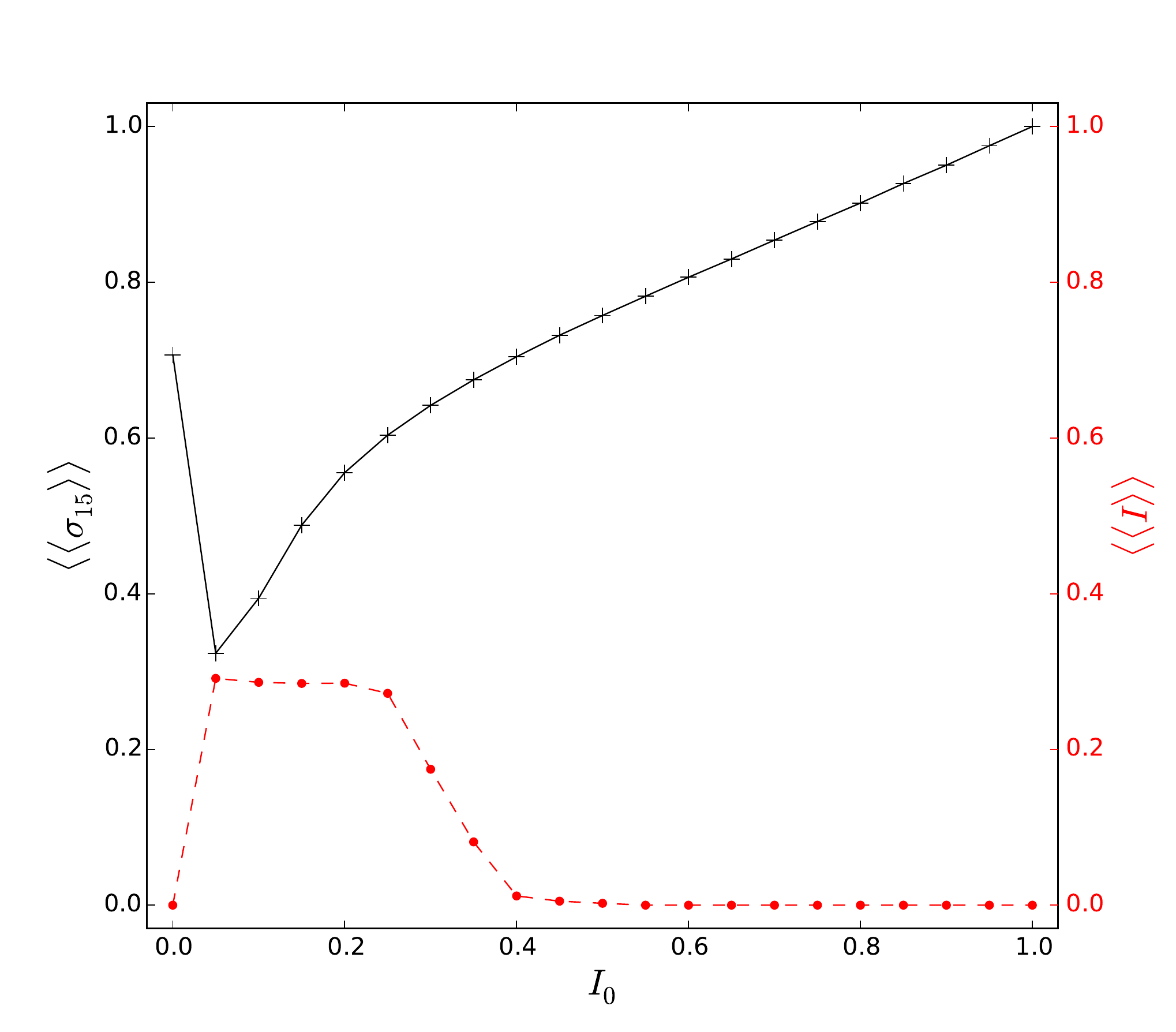}
\includegraphics[width=0.485\linewidth]{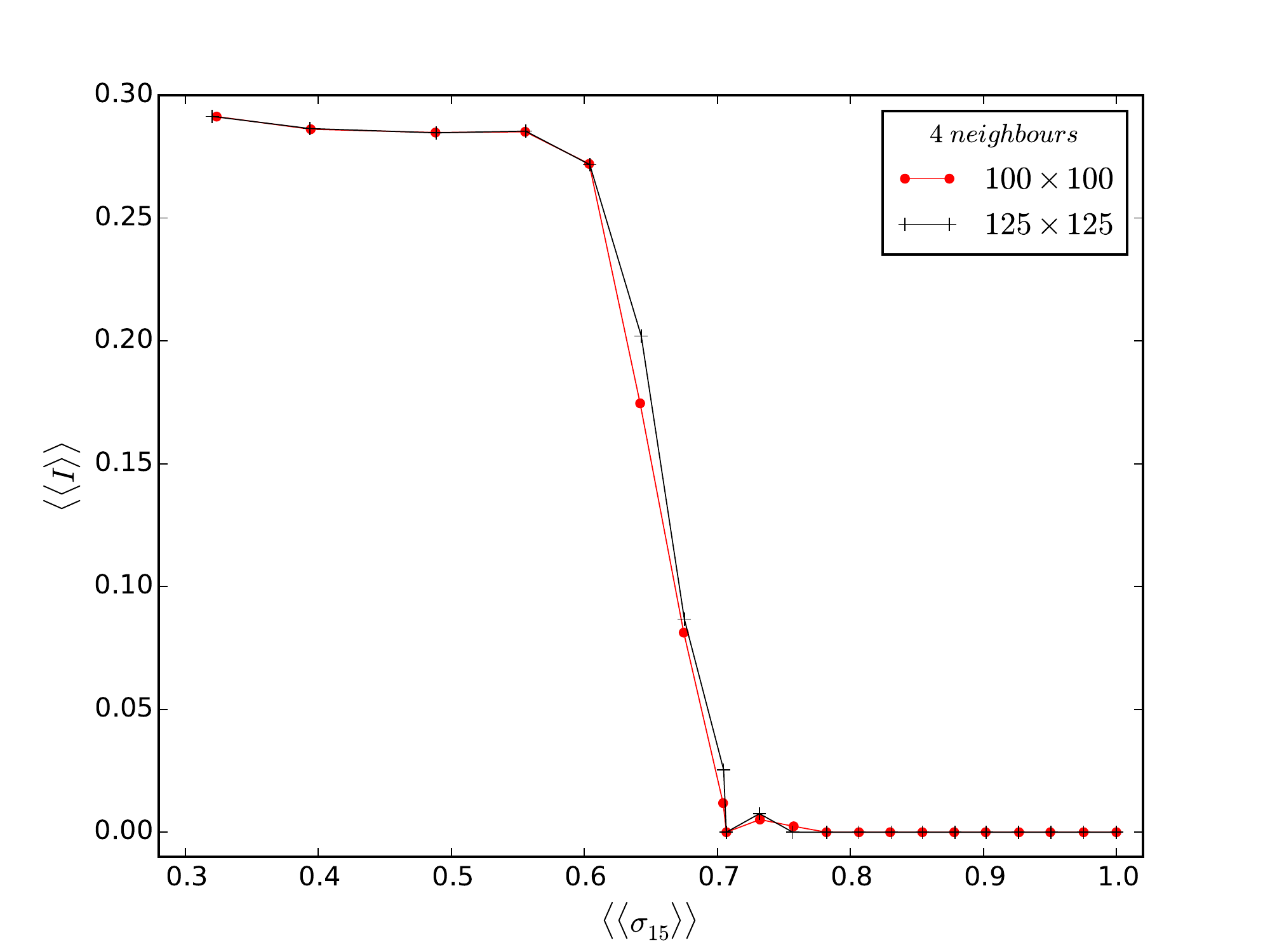}

\hspace{2cm} \scriptsize{(a)} \hfill \scriptsize{(b)} \hspace{2cm}
    \caption{\scriptsize{(a) Dependence of the transient synchronization order parameter $\langle \langle \sigma_{15} \rangle \rangle$ on the initial fraction of infecteds $I_0 \in [0,1]$ (with $S_0=R_0$ and system size $100 \times 100$). Here the synchronization order parameter at each $I_0$ is obtained by averaging over 100 random initial conditions. (b) Correlation between the asymptotic persistence parameter $\langle\langle I \rangle\rangle$ and the ensemble averaged transient synchronization order parameter $\langle\langle \sigma_{15} \rangle\rangle$. The quantities are obtained by averaging over $I_0 \in [0,1]$, with $S_0=R_0$. Here $K=4$.}}
    \label{sig15vsi0}
    \end{figure}
    
Specifically we will now investigate $\langle \langle\sigma_{15} \rangle \rangle$, namely the case where $T = \tau_D + 1$, where $\tau_D$ is the length of the disease cycle. So this quantity reflects the synchronization of the individual disease cycles in the population at early times, and can serve as an useful {\em order parameter for transient synchronization}. When $\langle \langle \sigma_{15} \rangle \rangle \rightarrow 1$, complete synchronization of the individual disease cycles in the population is obtained soon after one disease cycle.

Fig.~\ref{sig15vsi0}a shows the dependence of the degree of transient synchronization $\langle \langle \sigma_{15} \rangle \rangle$, namely the degree of synchronization right after completion of the first cycle of disease, on the fraction of infecteds $I_0$ in the initial population. It is evident that the onset of the persistence window is clearly indicated by minimum $\langle \langle \sigma_{15} \rangle \rangle$, i.e. the transient synchronization is the lowest when persist infection begins to emerge in the population. So, the {\em early synchronization properties of the system allows one to gauge the future persistence of contagion}. A valuable consequence of this observation is that early-time synchronization can serve as an early warning signal for sustained infection at a much later time.

Now we examine the explicit correlation between $\langle \langle \sigma_{15} \rangle \rangle$ and the asymptotic fraction of infecteds in the population $\langle\langle I \rangle\rangle$. This is shown in Fig.~\ref{sig15vsi0}(b), from where one can clearly see a well-defined transition to long-term persistent infection as the transient states get more synchronized. So the asymptotic fraction of infecteds decreases sharply at short-time synchronization order parameter values close to $2/3$. Namely, there exists a critical transient synchronization order parameter $\sigma^{\star}_T$, beyond which persistent infection does not occur (i.e. $\langle \langle I \rangle \rangle \sim 0$). Note that this critical $\sigma^{\star}_T$ reflects early-time properties, while offering a clear correlation with an asymptotic phenomena. It quantitatively confirms our intuition that when the system is more synchronous at early times, there is greater propensity of the infection dying out. 
  
So we conclude that greater degree of synchronization at early times hinders the sustenance of infection. Thus {\em early short-time asynchrony appears to be a consistent precursor to future persistence of infection, and can perhaps provide valuable early warning signals for anticipating sustained contagion in a population patch}.\\

\noindent
{\bf Transient Local synchronization:}
Now we explore the correlation of transient {\em local synchronization}, namely synchronization in a local neighbourhood of an individual. This is important, as infection spread is a local contact process and so the composition of its local nighbourhood is most crucial for an individual. In order to capture finite-time local synchrony, we introduce the following synchronization parameter:
\begin{equation}
\label{eqn:sigma}
\sigma_K^{(i,j)} (t) = | \frac{1}{K+1} \Sigma_{m,n} \exp^{i \phi_{m,n} (t)} |
\end{equation}
where $\phi_{m,n}$ is a geometrical phase corresponding to the disease stage $\tau_{m,n}$ of the individual at site $(m,n)$. Here the indices $m$ and $n$ run over the site index and all $K$ sites contained within the neighbourhood of $(i,j)$. The average of $\sigma_K^{(i,j)}(t)$ over all sites $(i,j)$ in the system is denoted by $\sigma_K (t)$.

The focus of our investigation is the finite time average of $\sigma_K(t)$ from initial time ($t=0$) to time $T$, where $T$ is of the order of one disease cycle length. We denote this measure of finite-time local synchronization as $\langle \sigma_{K,T} \rangle$. The ensemble averaged $\langle \sigma_{K,T} \rangle$ is denoted by $\langle \langle \sigma_{K,T} \rangle \rangle$, and this quantity reflects the typical transient local synchronization present in the system.

    \begin{figure}[H]
\centering
\includegraphics[height=0.475\linewidth,width=0.8\linewidth]{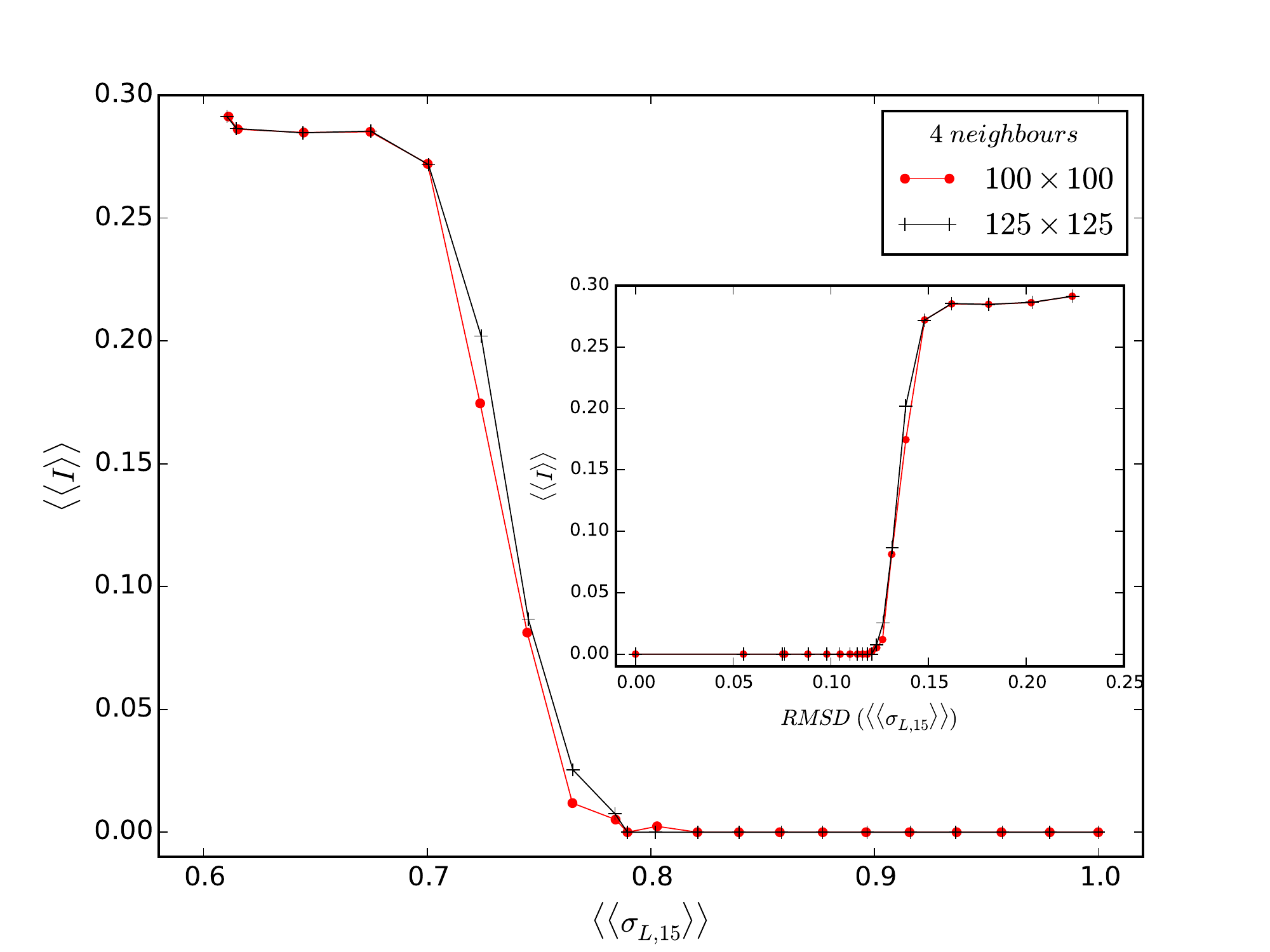}
\caption{\scriptsize{Dependence of the asymptotic persistence parameter $\langle\langle I \rangle\rangle$ on the ensemble averaged transient local synchronization order parameter $\langle\langle \sigma_{K,15} \rangle\rangle$. The quantities are obtained by averaging over $I_0 \in [0,1]$, with $S_0 = R_0$ and $K=4$. Inset shows the dependence of $\langle\langle I \rangle\rangle$ on the root mean square deviation (RMSD) of $\sigma_{K,15}$.}}
\label{localsync}
    \end{figure}
 
We show the explicit correlation between the transient local synchronization order parameter $\langle \langle \sigma_{K,15} \rangle \rangle$ and the asymptotic fraction of infecteds in the population $\langle\langle I \rangle\rangle$ in Fig.~\ref{localsync}. It is clearly evident that there exists a sharp transition to infection burn-out as transient local synchronization goes beyond a critical value $\sigma^{\star}_{K,T} \sim 3/4$. Broadly speaking then, on an average, local neighbourhoods need at least one neighbour whose state in not in sync with the other neighbours, to allow the sustenance of infection in the population. That is, when {\em local neighbourhoods are synchronized beyond a critical degree during early stage of disease spreading, persistent infection does not occur}. So, though critical $\sigma^{\star}_{K,T}$ depends on early-time spatially local information, it offers a clear indication of asymptotic phenomena. 

Further notice that the spread in transient local synchronization across initial states, as reflected by the root mean square deviation (RMSD) of $\langle \sigma_{K,T} \rangle$ in the inset of Fig.~\ref{localsync}, also exhibits a sharp transition from the case of non-persistent infection (i.e. $\langle\langle I \rangle\rangle = 0$) to persistent infection (where $\langle\langle I \rangle\rangle \sim 1/3$).  This suggests that when the early-time local synchronization has deviations larger than a critical RMSD, infection persists over a long time in the population patch.
 
\begin{figure}[H]
\centering
\includegraphics[height=0.475\linewidth,width=0.8\linewidth]{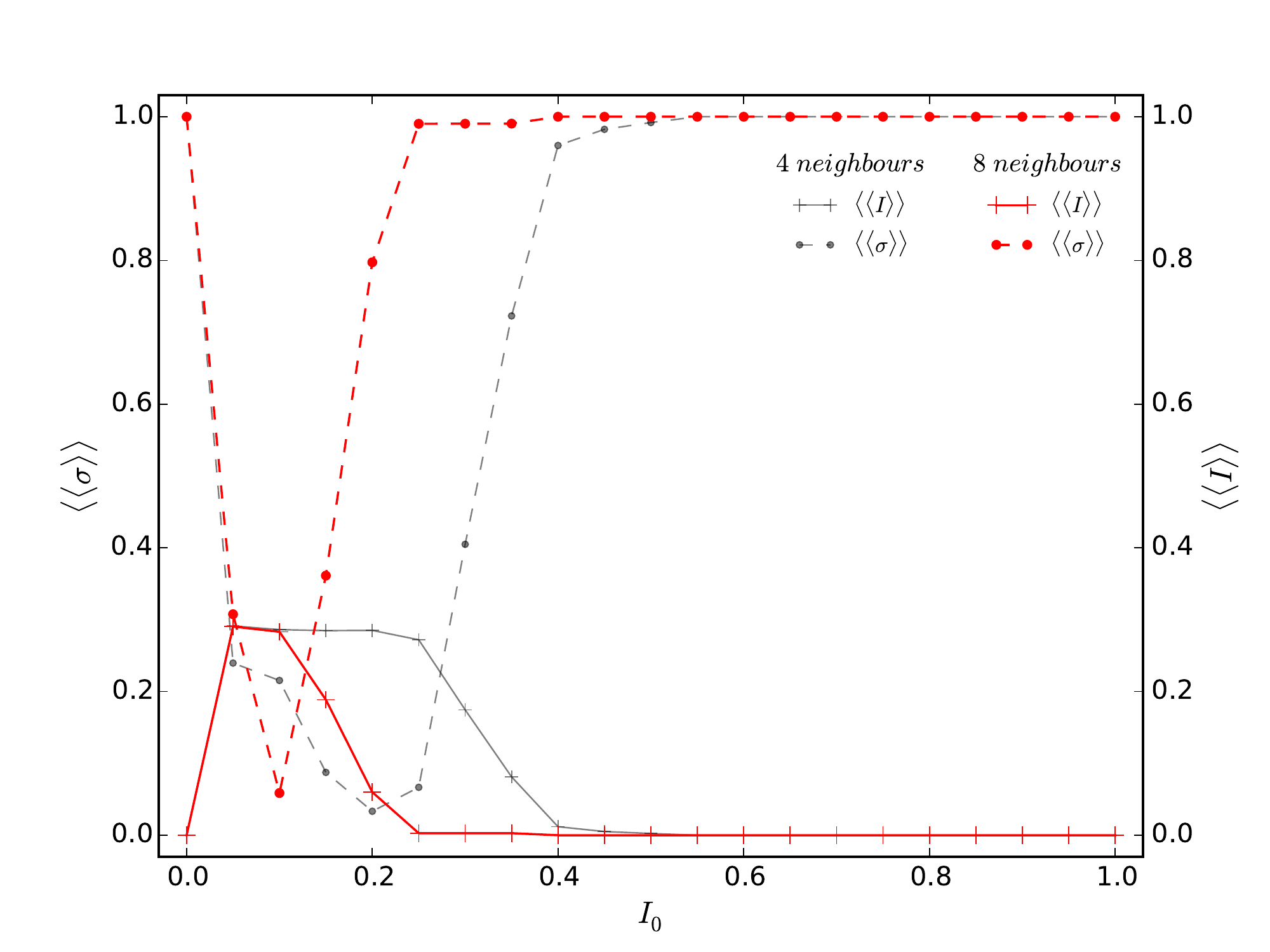}
\caption{\scriptsize{Dependence of $\langle \langle I \rangle \rangle$ on the initial fraction of infected individuals $I_{0}$ in the population, for the case of $K=8$ (solid red line) and the case of $K=4$ (red dashed line) for reference. Here system size is $100 \times 100$. Alongside we show the dependence of the ensemble averaged asymptotic synchronization order parameter $\langle \langle \sigma \rangle \rangle$  on $I_0$, for the case of $K=8$ (solid black line) and the case of $K=4$ (black dashed line) for reference.}}
\label{k8}
    \end{figure}

\noindent    
{\bf Dependence of persistence of infection on the range of infection transmission:}
Lastly, we explore the influence of the range of infection transmission on the persistence of infection. Specifically we investigate the case of $K=8$, namely the case where infected individuals can affect eight neighbours. So now the range of influence of the infected individual is double that presented earlier, where $K$ was $4$. Fig.~\ref{k8} shows the dependence of the persistence order parameter $\langle \langle I \rangle \rangle$ on the fraction of infected individuals $I_{0}$ in the initial population, with $S_{0} = R_{0}$. It is clearly evident from the figure that persistent infection is lower when the infected individual influences a larger number of neighbouring individuals. That is, surprisingly, a {\em larger range of infection transmission hinders long-term persistence of the disease.}

However, this counter-intuitive result is completely in accordance with our earlier observation, namely  higher synchronization implies lower persistence of infection. This is clearly bourne out by the asymptotic synchronization order parameter, which is also displayed in Fig.~\ref{k8} alongside the persistence order parameter $\langle \langle I \rangle \rangle$. From the figure it can be seen that for $K=8$ the synchronization is enhanced, and so $\langle \langle \sigma \rangle \rangle$ is low only in a very small range of $I_0$. It is this precise range that supports persistent infection. Since the range of low synchronization is significantly smaller for $K=8$ vis-a-vis $K=4$, we correspondingly have a significantly smaller range of persistent infection when the range of infection transmission is larger.

\begin{figure}[t]
\includegraphics[width=0.475\linewidth]{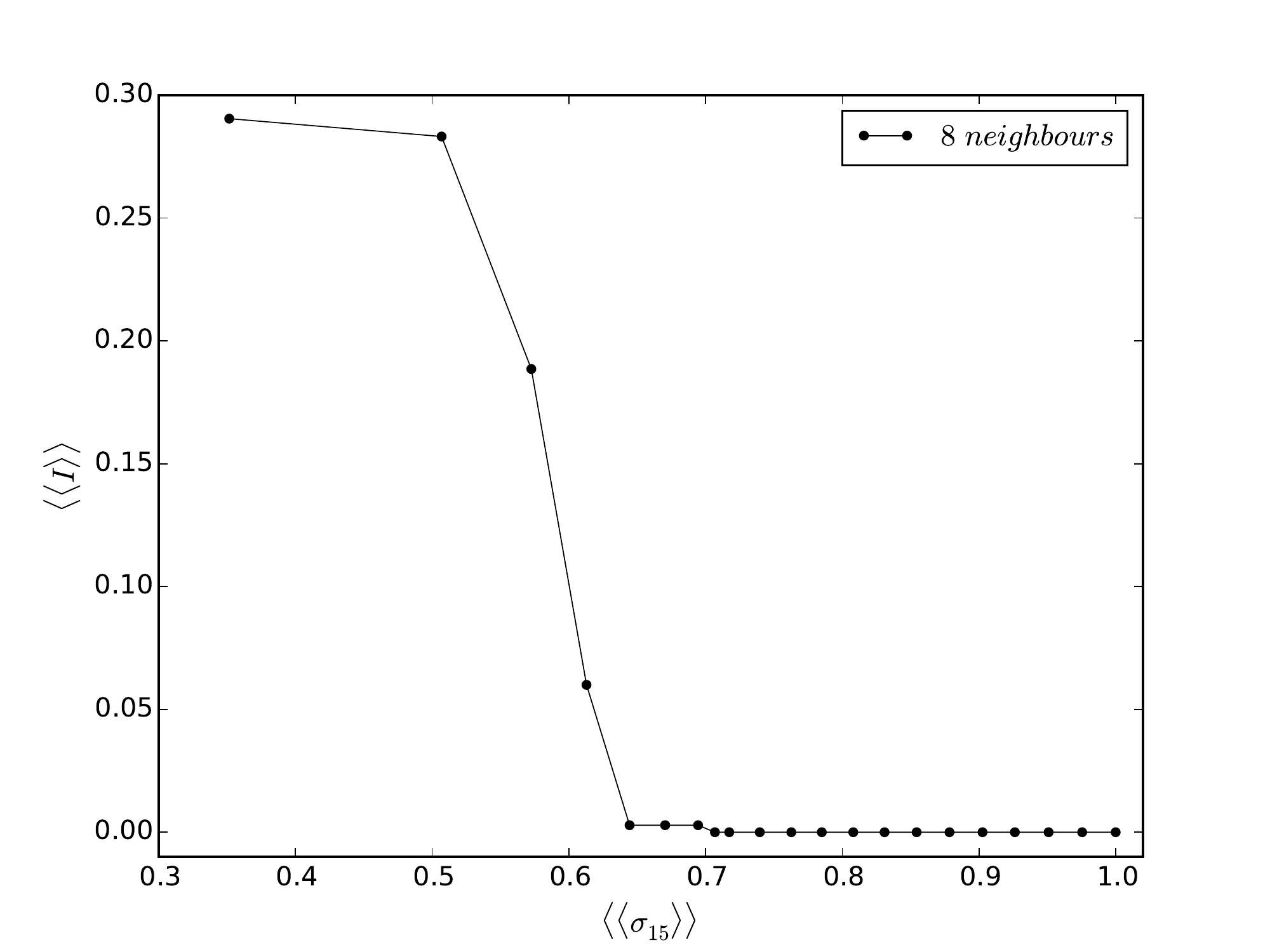}
\includegraphics[width=0.475\linewidth]{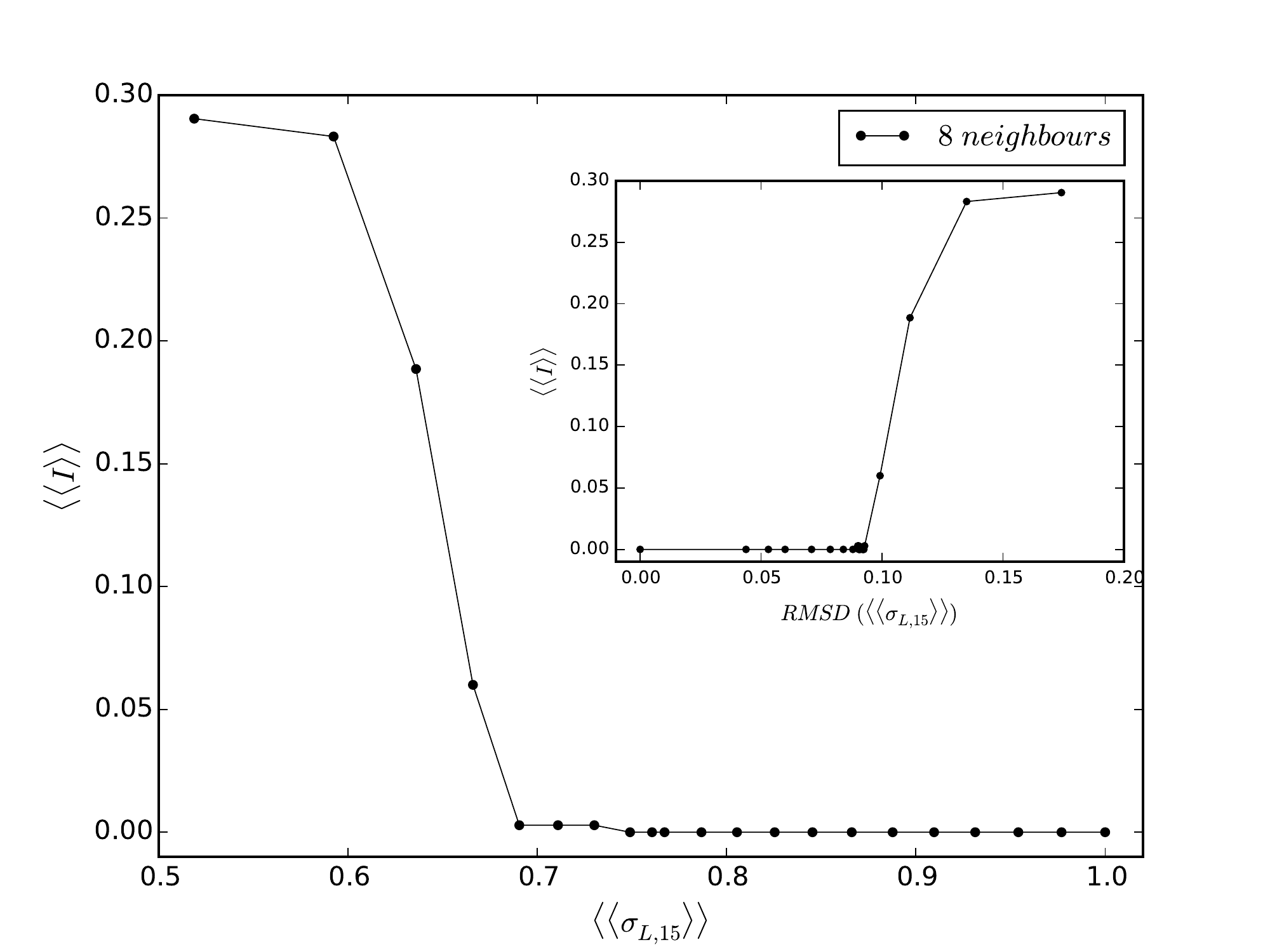}
\caption{\scriptsize{Dependence of the asymptotic persistence parameter $\langle\langle I \rangle\rangle$ on (left) ensemble averaged transient synchronization order parameter $\langle\langle \sigma_{15} \rangle\rangle$ and (right) the ensemble averaged transient local synchronization order parameter $\langle\langle \sigma_{K,15} \rangle\rangle$. The quantities are obtained by averaging over $I_0 \in [0,1]$, with $S_0=R_0$. Here system size is $100 \times 100$ and $K=8$.}}
        \label{fig3}
        \end{figure}
        
Further, we again examine the explicit correlation between the transient synchronization, as reflected by $\langle \langle \sigma_{15} \rangle \rangle$, as well as the local transient synchronization, as reflected by $\langle \langle \sigma_{K,15} \rangle \rangle$, and the asymptotic fraction of infecteds in the population $\langle\langle I \rangle\rangle$. These are shown in Figs.~\ref{fig3}a-b, from where one can again clearly see a well-defined transition to long-term persistent infection as the transient states get more synchronized both locally and globally. So again, quantitatively it can be seen that early-time local and global properties offer a clear indication of asymptotic persistence properties. This lends further credence to our central observation, and demonstrates the robustness and generality of the phenomenon with increasing range of infection transmission.

Also interestingly, as in the case of $K=4$, the asymptotic fraction of infecteds again decreases sharply at transient synchronization order parameter values close to $2/3$ and local transient synchronization order parameter values around $3/4$. However we observe that the precise value of the critical transient synchronization order parameters, $\sigma_T^{\star}$ and $\sigma_{K,T}^{\star}$, beyond which persistent infection does not occur (i.e. $\langle \langle I \rangle \rangle \sim 0$), is lower for the system with a wider range of infection transmission. This implies that larger de-synchronization of the phase of the individual disease cycles is necessary in order to obtain persistent infection, when the range of infection transmission is larger. This is consistent with the counter-intuitive observation that persist infection arises over smaller parameter ranges for larger $K$, as evident in Fig.~\ref{k8}.\\

\noindent  
{\bf Discussion}:
In summary, we have explored the emergence of persistent infection in a closed region where the disease progression of the individuals is given by the SIRS model, with an individual becoming infected on contact with another infected individual within a given range. We focussed on the role of synchronization in the persistence of contagion. Our key result is that higher degree of synchronization, both globally in the population and locally in the neighborhoods, hinders persistence of infection. Importantly, we found that {\em early local asynchrony appears to be a consistent precursor to future persistence of infection}, and can potentially provide valuable early warnings for sustained contagion in a population patch. Thus transient local synchronization can help anticipate the long-term persistence of infection. Further we demonstrated that when the range of influence of an infected individual is wider, one obtains lower persistent infection. This counter-intuitive observation can also be understood through the relation of synchronization to infection burn-out.

Lastly, our results also have broad relevance in the context of large interactive excitable systems. For instance, the system we study here is reminiscent of models of reaction-diffusion systems \cite{glass}, heterogeneous cardiac  tissue \cite{atr-fib} and coupled neurons \cite{neural}. The self-sustained excitations in these systems are analogous to the state of persistent infection we have focused on in this work. Specifically, persistent chaotic activity in a patch of tissue is characteristic of atrial fibrillation, and so our observations may have potential relevance to such phenomena arising in cardiac tissue. In the context of brain functions, neuronal circuits are able to sustain persistent activity after transient inputs, and studies have suggested that the asynchronous phase of synaptic transmission plays a vital role in the this persistent activity which is of considerable importance to motor planning and memory. Further, in the context of metapopulations \cite{eco}, there exists research which argues that enhanced coherence would decrease the probability of species persistence \cite{metapop}. So our demonstration of the potential of {\em early short-time local and global synchronization} as an early warning signal for anticipating persistent activity, has relevance to such phenomena as well.

\end{document}